\documentclass[sigconf, nonacm]{acmart}

\AtBeginDocument{%
  \providecommand\BibTeX{{%
    \normalfont B\kern-0.5em{\scshape i\kern-0.25em b}\kern-0.8em\TeX}}}

\usepackage{amsmath}
\usepackage[linesnumbered,lined,boxed,commentsnumbered]{algorithm2e}
\usepackage{graphicx}

\usepackage{bm}
\usepackage{listings}

\usepackage{xcolor}

\usepackage{xparse}

\NewDocumentCommand{\codeword}{v}{%
\texttt{\textcolor{blue}{#1}}%
}

\begin{document}

\title{Machine Learning Framework for Audio-Based Content Evaluation using MFCC, Chroma, Spectral Contrast, and Temporal Feature Engineering}

\author{Aris J. Aristorenas}
\affiliation{%
  \institution{Stanford University}
  \streetaddress{Department of Computer Science}
  \city{Stanford, CA}
  \country{USA} }
\email{aaris824@stanford.edu}

\begin{abstract}
\noindent  {\it This study presents a machine learning framework for assessing similarity between audio content and predicting sentiment score. We construct a dataset containing audio samples from music covers on YouTube along with the audio of the original song, and sentiment scores derived from user comments, serving as proxy labels for content quality. Our approach involves extensive pre-processing, segmenting audio signals into 30-second windows, and extracting high-dimensional feature representations through Mel-Frequency Cepstral Coefficients (MFCC), Chroma, Spectral Contrast, and Temporal characteristics. Leveraging these features, we train regression models to predict sentiment scores on a 0-100 scale, achieving root mean square error (RMSE) values of 3.420, 5.482, 2.783, and 4.212, respectively. Improvements over a baseline model based on absolute difference metrics are observed. These results demonstrate the potential of machine learning to capture sentiment and similarity in audio, offering an adaptable framework for AI applications in media analysis.} \newline
\end{abstract}


\maketitle

\section{Introduction}

In today’s digital landscape, social media platforms like TikTok, Instagram, and YouTube have become essential for artists and musicians to share their performances and gain audience feedback. When creators upload content, they are unable to revise it post-publication without losing valuable metrics such as likes and comments. This limitation is especially challenging for musicians who upload song covers, as it restricts their ability to make quality improvements based on initial audience reactions. Traditional quality checks, such as previewing content for test audiences, are costly and time-intensive, making them unfeasible for many creators.

This study proposes a machine learning model to perform this quality check in a way that is automated, repeatable, and does not require any external human input (like a test audience). The model works by providing creators with a predictive sentiment score. We train our model using audio samples from covers and original songs along with sentiment scores aggregated from social media comments. By analyzing these inputs, the model generates a predicted sentiment score on a 0-100 scale, representing anticipated audience reception if the cover were published.

The innovation in this study lies in the tailored model architecture and feature engineering, which goes beyond simple audio comparisons. Our baseline approach—calculating absolute differences between cover and original—fails to capture the nuanced changes in tone, style, and instrumentation that often resonate positively with audiences. Instead, we use advanced feature representations like Mel-frequency Cepstral coefficients (MFCC) for timbral nuances, along with Chroma, Spectral Contrast, and Temporal features, to train regression models that more accurately predict audience sentiment.

\section{Related work}

Slaney et al. (2008) proposed a technique for embedding songs into Euclidean space to develop a metric for semantic dissimilarity, drawing a parallel to cosine similarity in word embeddings to measure similarity between words. They applied algorithms like kNN and LDA for song classification \cite{a1}. Dhanaraj et al. (2005) explored hit song prediction using a dataset of 1,700 songs, applying acoustic and lyric features with classifiers such as SVM and boosting. They found that lyric features best predicted a song’s success, while combining these with acoustic features offered minimal improvement \cite{a2}.  Both Chang et al. (2017) and Mehta (2019) developed CNN-based models to identify covers, using a cross-similarity matrix to detect learnable patterns between cover and original songs \cite{a3, a4}. Another related work on song covers is by Osmalsky (2015) who approached the problem of song cover identification by combining features \cite{a5}. The authors trained classifiers using features such as global tempo, duration, loudness, beats, and chroma average, and aggregated the different classifiers into combinations of hybrid classifiers which use one or more features. It was found that the hybrid classifiers outperformed the single classifiers.

Ravuri et al. (2010) also focused on cover detection, applying SVM and multi-layer perceptron models with normalization processing, achieving an error rate of 10\%—an improvement over the then state-of-the-art of 21.3\% \cite{a6}. Silva, et al. (2015) proposed a method based on time-series shapelets to also address the cover identification problem. Their method differs from the previous works, as it did not require performing feature extraction on the entire song. Instead, they train a model that learns phase features of original songs based on small excerpts of it, and they use these small segments to search for song covers \cite{a7}. One motivation for doing so is for copyright (sometimes covers cannot be created because an original song has some copyright protection). Silva-Reyes (2017) proposed features such as pitch, timbre, and beat information for cover identification, noting that frequency processing with beat alignment performed comparably to the state-of-the-art \cite{a8}.

Fang et al. (2017) incorporated chroma features with dimensionality reduction and an autoencoder to streamline chroma-based pairwise comparisons \cite{a9}. The rationale of using dimensionality reduction was that traditional chroma-based pairwise-comparisons between songs was expensive. Huang et al. (2016) applied the 2D Fourier Transform to project songs into vector space \cite{a10}. By doing so, they could leverage machine learning classifiers like the nearest neighbor algorithm to detect song similarity. Correya et al. (2018) introduced a model that incorporates lyrics and metadata (like song titles) for cover detection. It was found that using TF-IDF alone for text similarity between songs led to a 35.5\% increase in the classifier's precision for cover identification \cite{a11}.  This is plausible for cover detection because covers will typically have lyrics that match the original song. Finally, Kim et al. (2008) made use of a chroma-based dynamic feature vector. The authors claim that this feature vector describes pitch changes in a song. The motivation for using this feature vector for the classifier is that humans identify songs in a similar way according to the human auditory system \cite{a12}.

\section{Dataset and Features}

Custom data collection efforts were required in order to train the proposed model. This section discusses the data collection steps, as well as the feature engineering steps.

\textbf{Data collection}: The goal of data collection is to retrieve video covers of songs, as well as the original version of the song on social media platforms which allow users to comment on the cover. The primary challenge was to design a way to automate the collection. This was broken down into three main tasks: 1) automating the search of YouTube covers, and the original song version corresponding to the cover, 2) automating the retrieval of the music cover, and original song, 3) automating the retrieval of the comments pertaining to the cover.  YouTube was selected to support the automation due to the availability of a public API, which allows for the programmatic retrieval of both YouTube comments, and the video for some given video ID.

A script was developed to process a list of original songs formatted as "<Song Title> - <Artist>." The script utilizes the YouTube API, passing in the list along with an API key to retrieve video IDs and associated metadata, including the upload date, view count, like count, comment count, and the number of subscribers on the channel. An example of the data frame is shown in Table ~\ref{tab:video_metadata} for three retrieved covers of a popular song by a well-known artist in the U.S.

\begin{table}[ht]
\centering
\caption{Example metadata for retrieved video covers}
\label{tab:video_metadata}
\begin{tabular}{rllrr}
  \hline
 & video\_id & upload\_date & views & likes \\ 
  \hline
1 & Duaxp1nc5po & 2016-11-03T18:30:00Z & 6342336 & 113146 \\ 
2 & TnjS5kqyu3E & 2022-11-02T01:05:46Z & 1129167 & 22417 \\ 
3 & jw3MqySX9qw & 2021-10-01T10:00:23Z & 1603208 & 48048 \\ 
   \hline
\end{tabular}
\end{table}

With these video IDs available, the next step was to programmatically download the video files in .mp4 format, along with the video's comments. An example data frame showing the comments for one musician's cover is provided in Table ~\ref{tab:example_comments}.

\begin{table}[ht]
\centering
\caption{Example comments for a retrieved video cover}
\label{tab:example_comments}
\begin{tabular}{rl}
  \hline
& comment \\ 
  \hline
1 & I like you voice, keep it going. I will go on alive a long.... \\ 
2 & You are wonderful!!! I love this song. Performance..the best!!! \\ 
   \hline
\end{tabular}
\end{table}

\textbf{Sentiment score}: The pre-processing step for the retrieved YouTube comments was to calculate the sentiment score. To train a predictive model using audio as input, the NLTK library was used to assign a sentiment score to each comment on a 0-100 scale, where 0 indicates the most negative sentiment and 100 the most positive. These scores served as the labels for the training dataset. The results are shown in Table ~\ref{tab:sentiment_scores}.

\begin{table}[ht]
\centering
\caption{Example comments and their calculated sentiment scores}
\label{tab:sentiment_scores}
\begin{tabular}{rllr}
  \hline
 & comment & video.ID & sentiment.score \\ 
  \hline
1 &  I like your voice... & Duaxp1nc5po & 50.00 \\ 
2 & You are wonderful!!!... & Duaxp1nc5po & 96.78 \\ 
   \hline
\end{tabular}
\end{table}

\textbf{Audio waveforms}: In order to retrieve the audio data in WAV format, the videos from YouTube of each cover, and original were downloaded programatically using the YouTube API. From here, a script was written to extract the audio from the video. The purpose of downloading the video instead of just the audio using the API was for future work in which computer vision techniques could be used to extract features from the video to also help predict sentiment.


\textbf{Training examples}: The final steps in preparing the training examples involved: 1) calculating an aggregate sentiment score per video cover to form the labels, 2) using the \textbf{librosa} Python library to read the audio data into a vector, and 3) performing a join on all data frames by video ID. The resulting raw training dataset consisted of rows made up of two vectors $v_1, v_2 \in \mathbb{R}^\text{d}$: the audio data for the cover, and the audio data for the original song, as well as the aggregated sentiment label, $y \in \mathbb{R}$. Table ~\ref{tab:training_examples} shows three rows of the training examples data frame.

\begin{table}[ht]
\centering
\caption{Example of audio data vectors and corresponding sentiment score}
\label{tab:training_examples}
\begin{tabular}{rllr}
  \hline
 & Audio Data\_cover & Audio Data\_original & avg sentiment \\ 
  \hline
1 & [0. 0. 0. ... 0. 0. 0.] & [... 0.0082 0.0047] & 66.99 \\ 
  2 &  [0. 0. 0. ... 0. 0. 0.] &  0.0082 0.00457 ...] & 67.12 \\ 
  3 & [0. 0. 0. ... 0. 0. 0.] & [... 0.0082 0.00457 ...] & 77.18 \\ 
   \hline
\end{tabular}
\end{table}

\textbf{Feature Engineering}: The raw training examples presented several challenges: 1) the covers varied substantially in audio length from each other and from the original, 2) the covers were often misaligned from the original song in terms of the starting beat, and 3) some covers included unrelated audio, such as spoken introductions, advertisements, or sponsorship messages. Additional acoustic challenges included: 1) significant stylistic differences from the original (for example, a rock rendition of a pop song), 2) variations in musical key between the cover and original, and 3) differences in instrumentation, such as a piano cover of a blues song. Data pre-processing and feature extraction techniques were applied to address these issues.

\textbf{Data pre-processing}: The main data pre-processing steps used involved re-sampling, padding, normalization, and segmentation. Re-sampling was necessary because the audio covers and originals differed in sampling rates. Since sampling rate was assumed not to affect sentiment prediction, all audio was re-sampled to a standard rate of 22,050 Hz to maintain consistency and prevent issues like time warping.

\textbf{Data augmentation}: In order to address the issue of differing audio lengths between a cover and the original version, segmentation was applied, where fixed portions (or windows) of the audio were extracted, re-sampled, and compared. For example, 30 second segments of the cover and the original were extracted, followed by resampling to a standard rate of 22050 Hz, to form a new training example set. This process acts as a feature map: $\phi (\textbf{x})$ where $\phi: \mathbb{R}^{d_1 + d_2} \to  \mathbb{R}^{d_1' + d_2'}$. As a result, each row in the new training dataset represents segments of audio, with each former cover-original pair yielding an additional 6-7 rows (on average) after segmentation, effectively augmenting the data.

\section{Methods}

We now formulate the learning algorithms used for each of the models. The goal here is to understand how to derive the loss function, and is demonstrated on the MFCC feature. A similar derivation can be shown for the Chroma, Spectral Contrast, and Temporal features.

\textbf{Predicting sentiment score:} The data collection and data pre-processing steps resulted in training examples of the form $(x^{(i)}, y^{(i)})$, where $x^{(i)} = [x_{cover}^{(i)}; x_{original}^{(i)}] \in \mathbb{R}^{d}$ and $y^{(i)} \in \mathbb{R}$ is the sentiment score. In this formulation, $x_{cover}^{(i)} \in \mathbb{R}^{d_1}$ represents the vector for the audio cover, and $x_{original}^{(i)} \in \mathbb{R}^{d_2}$ represents the vector for the original audio. The concatenated vector $x^{(i)}$ then combines information from both the cover and original audio.

The mathematical representation of the concatenated input vector is given by:

\begin{align}
x^{(i)} = [x_{cover}^{(i)}; x_{original}^{(i)}] \in \mathbb{R}^{d}
\end{align}

\noindent where $d = d_1 + d_2$. During segmentation, we extracted 30-second windows (the segments) of both the cover and original as follows: ${x_{seg}}^{(i)}_k = [{x_{cover}}^{(i)}_k; {x_{original}}^{(i)}_k] \quad \text{for } k = 1, 2, \ldots, N$ where:

\begin{align}
{x_{audio}}^{(i)}_k = x_{audio}^{(i)}[S_k : E_k]
\end{align}

\noindent and $S_k = (k-1) \times L + 1 \text{ and } E_k = \min(k \times L, \text{len}(x_{audio}^{(i)}))$. If $\text{len}({x_{audio}}^{(i)}_k) < L$, we pad using:

\begin{align}
{x_{audio}}^{(i)}_k = \text{pad}({x_{audio}}^{(i)}_k, L)
\end{align}

\noindent This effectively performs data augmentation, where the new training examples $({x_{seg}}^{(i)}_k, y^{(i)}_k)$ where ${x_{seg}}^{(i)}_k \in \mathbb{R}^{L\cdot SR}$ now reside in a higher-dimensional space due to augmentation. Here, $L$ is the length of each segment (in seconds), $SR$ is the sampling rate in Hz, and $k = 0, 1, 2, \ldots, |S|$ represents the number of segments for each original example.

 \textbf{MFCC Feature:} For each segmented audio ${x_{seg}}^{(i)}_k$, Mel-Frequency Cepstral Coefficients (MFCCs) are extracted to transform the raw audio data into a set of features that capture the timbral and textural aspects of the audio signal.

\textbf{Frame the Signal into Short Frames:} The audio segment is divided into short time frames to capture the time-varying nature of the audio signal. For a given frame length $F$ and hop length $H$, the $j$-th frame of the $k$-th segment for the $i$-th example is represented as:

\begin{align}
x_{frame}^{(i, k, j)} = {x_{seg}^{(i)}}_k[n: n + F] \quad \text{for } n = (j-1) \times H, j = 1, 2, \ldots, J
\end{align}

\noindent where $J$ is the total number of frames in the segment.

\textbf{Fourier Transform:} Each frame is transformed from the time domain into the frequency domain using the Fast Fourier Transform (FFT):
\begin{align}
X_{frame}^{(i, k, j)} = \text{FFT}(x_{frame}^{(i, k, j)})
\end{align}

\textbf{Mel Filterbank:} The power spectrum obtained from the FFT is passed through a Mel filterbank, which is a collection of filters that emulate the human ear's response to different frequencies. The output of the $m$-th Mel filter for the $j$-th frame is given by:
\begin{align}
M_{frame}^{(i, k, j, m)} = \sum_{f} |X_{frame}^{(i, k, j)}(f)|^2 \cdot H_m(f) \quad \text{for } m = 1, 2, \ldots, M
\end{align}

\noindent where $H_m(f)$ is the $m$-th Mel filter and $M$ is the total number of Mel filters.

\textbf{Logarithmic Scaling:} The energies of the Mel filterbank are scaled logarithmically to better match human perception:
\begin{align}
L_{frame}^{(i, k, j, m)} = \log(M_{frame}^{(i, k, j, m)})
\end{align}

 \textbf{Discrete Cosine Transform (DCT):} The log Mel filterbank energies are then transformed using the Discrete Cosine Transform to obtain the MFCCs:
\begin{align}
C_{frame}^{(i, k, j)} = \text{DCT}(L_{frame}^{(i, k, j)})
\end{align}

\noindent The final MFCCs for each segment are typically computed by averaging across all frames within the segment:

\begin{align}
\text{MFCC}^{(i, k)} = \frac{1}{J} \sum{j=1}^{J} C_{frame}^{(i, k, j)} = \phi({x_{seg}}^{(i, k)}) 
\end{align}

\noindent The result is a fixed-size vector of MFCCs for each audio segment in the cover, and original song which aim to capture the essential characteristics for the machine learning regression task. This is precisely the feature map we sought: $\phi({x_{seg}}^{(i, k)})$, where $\phi: \mathbb{R}^{L \cdot SR} \to \mathbb{R}^M$ and where $M$ is the number of MFCC coefficients derived from the Mel filterbank's output, and $J$ is the total number of frames in the segment. This transformation encapsulates the essential timbral characteristics of the audio segment into a compact, fixed-size feature vector.

\textbf{Loss function:} For the regression task of predicting sentiment score, each model developed aims to minimize the mean-squared error (MSE) loss. The MSE loss function is given by:

 \begin{align}
 J(\theta) = \frac{1}{N} \sum_{i=1}^{N} (y^{(i)} - \hat{y}^{(i)})^2
 \end{align}
 
 \noindent where $J(\theta)$ represents the MSE loss function, $y^{(i)}$ is the actual sentiment score for the $i$-th example, $\hat{y}^{(i)}$ is the predicted sentiment score by the model for the $i$-th example, $\theta$ represents the model parameters, and $N$ is the total number of examples. The model is given by:

\begin{align}
\hat{y}^{(i)} = \theta^T \phi({x_{seg}}^{(i)})
\end{align}

\noindent where $\phi$ is again the feature map that transforms each audio segment into a feature vector representing its feature (like MFCCs).


\section{Experiments, results, discussion}

The main form of experimentation of this study involved trying to extract features which resulted in better model performance in terms of RMSE over the baseline.


\textbf{MFCC model:} A plot of the extracted MFCCs is shown below in Figure ~\ref{fig1} for the first 3 segments of each cover, and original pair.


\begin{figure}[H]
  \centering
  \frame{\includegraphics[width=\linewidth]{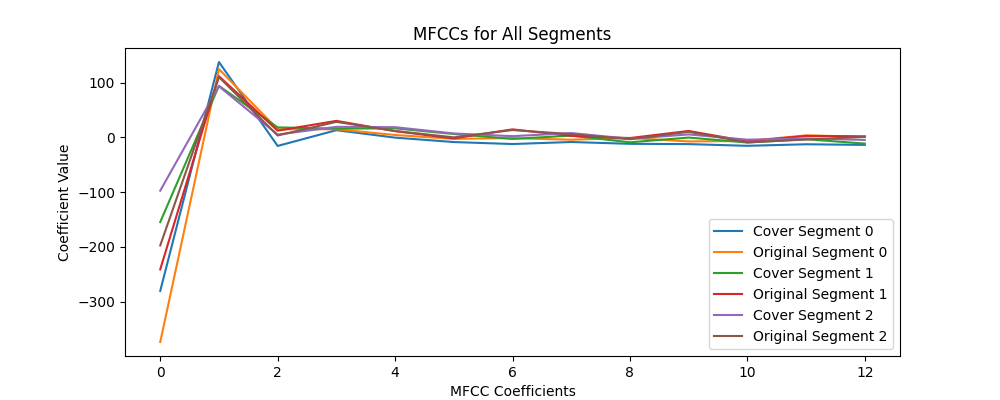}}
  \caption{MSE loss for the MFCC model}
      \label{fig1}
\end{figure}

\begin{figure}[H]
  \centering
  \frame{\includegraphics[width=\linewidth]{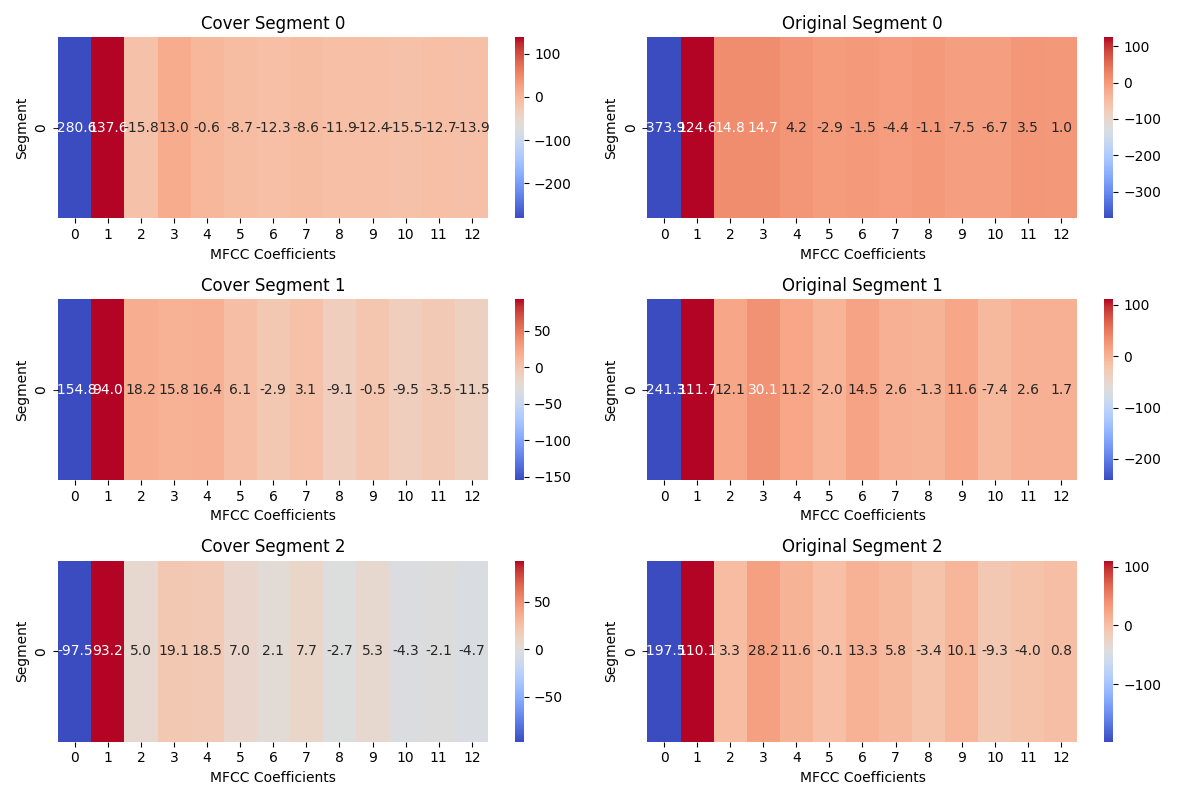}}
  \caption{MSE loss for the MFCC model}
      \label{fig-heat-map}
\end{figure}

The x-axis represents the index of the Mel-Frequency Cepstral Coefficients (MFCCs). These coefficients are features derived from the audio signal that capture the power spectrum of the sound. Typically, the first few coefficients (low index values) capture the most significant characteristics of the audio signal, such as the general shape of the spectral envelope. This explains the initial spike at MFCC = 1 (the first MFCC coefficient), and is common since the first MFCC coefficient represents the general loudness or spectral energy level of the audio. The y-axis coefficient values are essentially a measure of the energy in different frequency bands of the audio signal. They are derived from the log power spectrum of the short-time Fourier transform of the audio signal, followed by a Mel-scale transformation and discrete cosine transform. The MFCCs (after Mel-scale transformation) reflect the overall energy distribution in various frequency bands as perceived by the human ear. The coefficients from MFCC 2 to 12, usually capture more subtle and detailed aspects of the sound, such as the texture and timbre. Stability in these coefficients suggests a consistent texture and timbre across the different segments of the cover and original tracks. Another way to visualize these features is using a heatmap, shown in Figure ~\ref{fig-heat-map}. This again shows that the concentration of energy at the early MFCC coefficients.


Since the MFCC patterns for both cover and original tracks are nearly identical with minimal deviation, this suggests a close match in their spectral properties, which is expected since the covers aim to be close to the original song. The small deviations between segments could be indicative of subtle differences in the style, instrumentation, or recording quality between the cover and the original. These differences might not be drastic enough to significantly affect the overall similarity in terms of the spectral profiles.

\begin{figure}[!htbp]
  \centering
  \frame{\includegraphics[width=\linewidth]{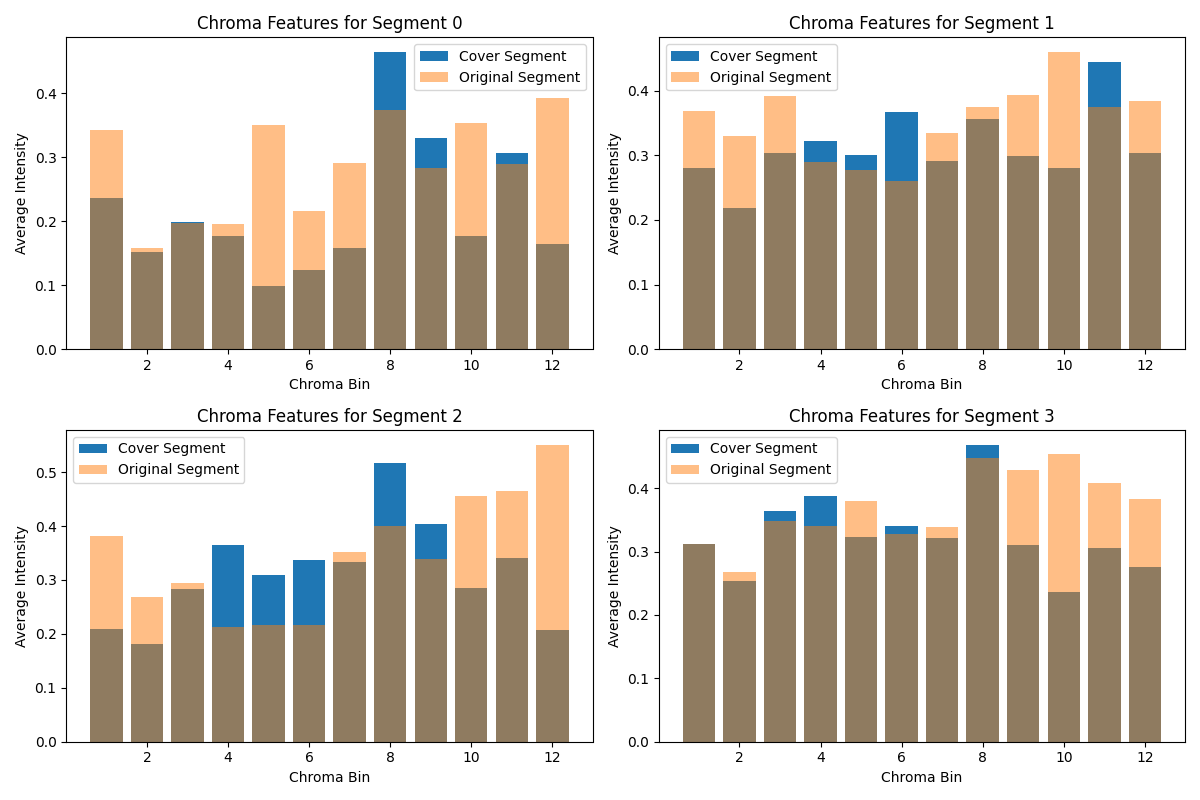}}
  \caption{Chroma features for the first 4 segments}
      \label{fig-chroma-features}
\end{figure}

\textbf{Chroma model:} The Chroma features plot is shown in Figure ~\ref{fig-chroma-features} for the first 4 segments of the cover and original pairs. These features represent a distribution of the musical energy across different pitch classes. The bins on the x-axis correspond to the 12 different pitch classes: C, C\#, D, D\#, E, F, F\#, G, G\#, A, A\#, B, which comprise of one full octave. The y-axis represents average intensity. The height of each bar indicates the average intensity (or presence) of each pitch class in the audio segment. Higher bars mean that the corresponding pitch class is more prominent in that segment. The orange bars represent the intensity of the original segments, while the blue bars represent the cover segment. The brown bar portions represent overlap to distinguish where the cover and original coincide and overlap. This overlap helps to visually identify similarities in pitch prominence between cover and original.


Chroma features capture the essence of the harmonic content of the music. By comparing the chroma features of the cover and the original song, we can assess how similar or different their harmonic structures are. The variation in each segment of the bar plots is desirable: we want to learn the weights during training which capture this variation as it relates to sentiment score. This is motivated by how the harmonic content of a song is closely related to its emotional and sentimental impact. In other words, the variation in Chroma features can help identify differences in vocal timbre, genre, and instrumentation.

\begin{figure}[H]
  \centering
  \frame{\includegraphics[width=\linewidth]{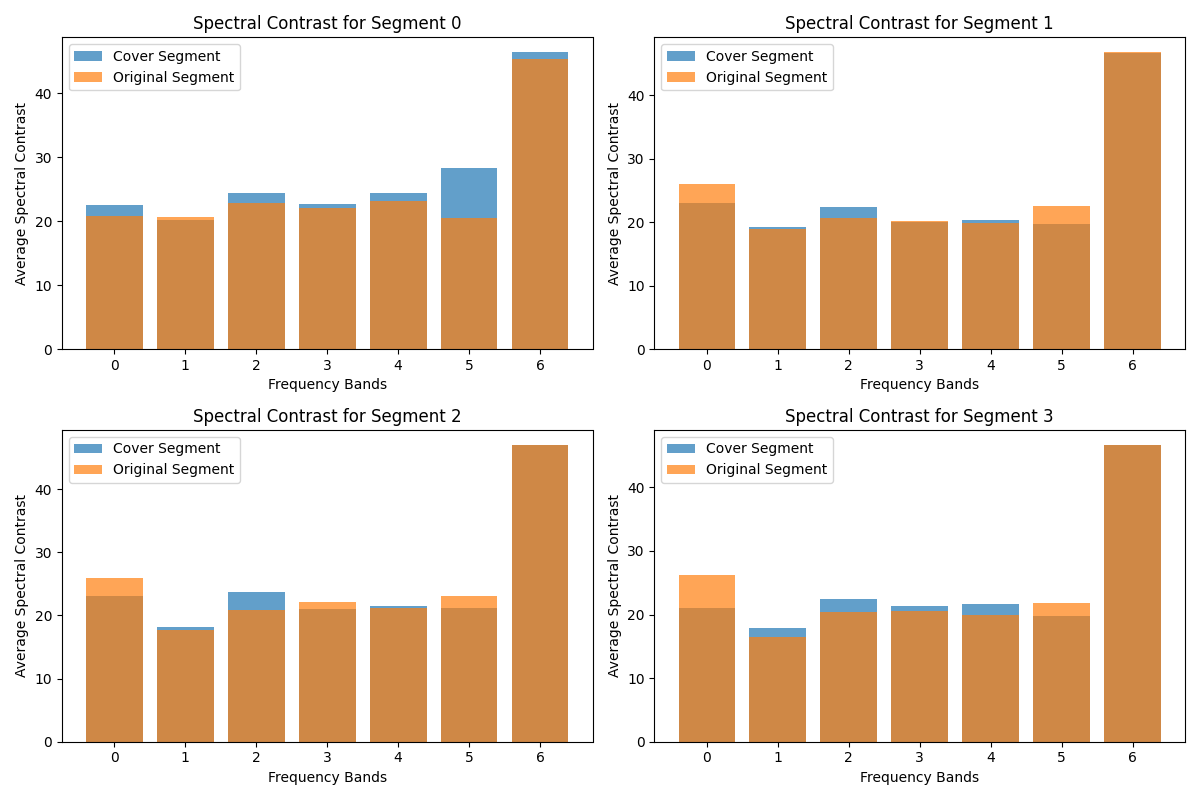}}
  \caption{Spectral contrast features for the first 4 segments}
      \label{fig-spectral-features}
\end{figure}

\textbf{Spectral Contrast model:} Another bar plot is shown in Figure ~\ref{fig-spectral-features} showing the average spectral contrast for each frequency band (for the first 4 segments). For bands 0 to 5, the average spectral contrast is consistently around 25. Furthermore, there are few differences between the cover version and original version. At frequency band 6, the average spectral contrast is nearly doubled to 50. This suggests that these frequency ranges have similar spectral textures - similar distributions of energy across frequency bands - in both versions of the songs. This can imply a close resemblance in the lower to mid-frequency range, which often captures the bulk of the musical content like melody and harmony. A significantly higher average spectral contrast in band 6 could be due to various reasons such as different instrumentation, recording quality, or artistic interpretation in the cover version.


\subsection{Discussion of results}

We now discuss in further detail the key findings of this study.

\textbf{Hyperparameters for training:} The training hyperparameters chosen resulted in convergence with no indication of overfitting, and had acceptable RMSE values. The choices involved: 1) using incremental learning with warm start for continuous model training, 2) a tolerance of 0.003, 3) learning rate of 0.01, and 4) an L2 regularization alpha value of 0.0001.

\begin{figure}[H]
  \centering
  \frame{\includegraphics[width=\linewidth]{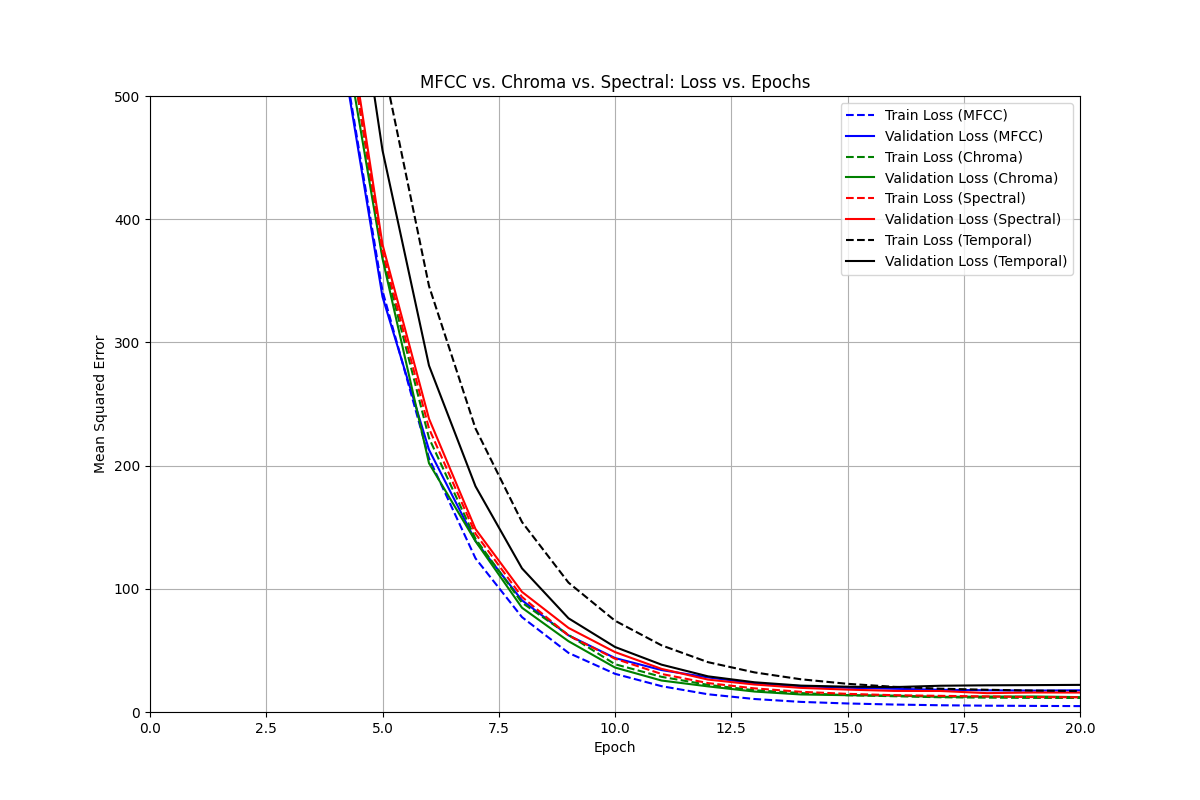}}
  \caption{MSE loss for the MFCC, Chroma, Spectral, and Temporal models}
      \label{fig-loss}
\end{figure}

\textbf{Learning rate loss curves:} The loss curves are shown in Figure ~\ref{fig-loss} for each of the models during training and validation.


\begin{figure}[H]
  \centering
  \frame{\includegraphics[width=0.8\linewidth]{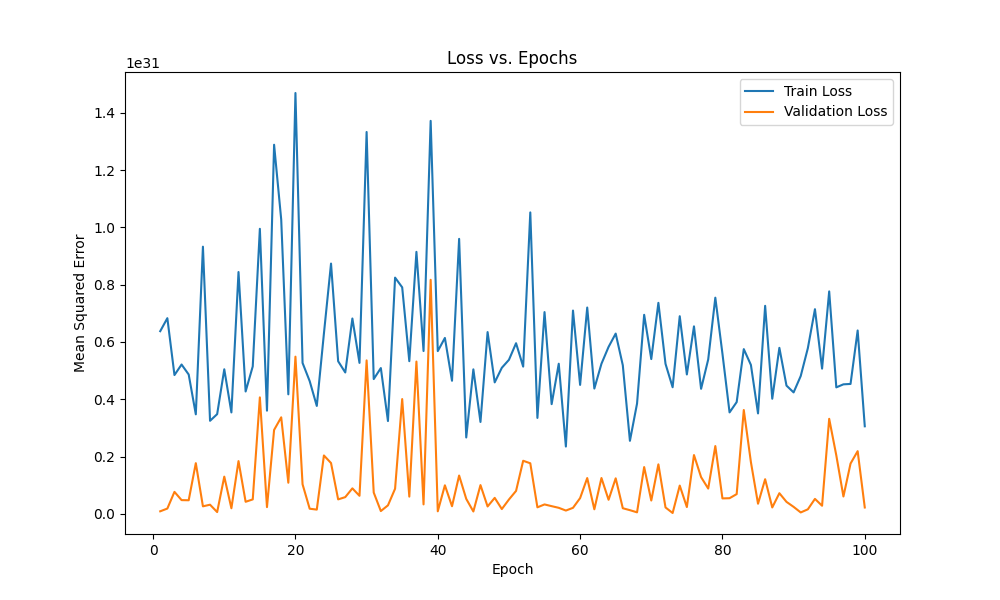}}
  \caption{MSE loss for the absolute difference model}
      \label{fig-baseline}
\end{figure}

All four models seem to perform acceptably well: the parameters are learned in early epochs and converge after approximately 10 epochs. The validation loss (dashed lines) also has a similar pattern indicating that overfitting is not an issue. Figure ~\ref{fig-baseline} shows the same loss curve for the baseline, made up of taking the absolute differences between cover, and original in the segmented data. As seen, the baseline model does not converge. This is a consequence of the feature choice, which underscores the importance of using more sophisticated features like MFCCs, Chroma, Spectral Contrast, and Temporal features.

This further supports the use of these specific features for the machine learning task at hand, which was to learn a model that could capture the nuances in musical interpretation. The covers that made up this data set involved different musical keys from the original, the use of different instruments, the use of different renditions (one involved a Reggaeton cover of a piano pop song).


\textbf{Root Mean Square Errors:} Table ~\ref{tab:rmse-models} summarizes the RMSE for all the ML models. As shown, the Spectral Contrast model performed the best. All the feature-based models performed over the baseline, demonstrating the advantage of using advanced audio features over absolute difference alone. Note that the RMSE value for the baseline model of 3.723e+14 is on a different scale than the other models, but nonetheless, is indicative of how the absolute difference approach does not effectively learn the task due to an out-of-range RMSE, especially when compared to the feature-based models.

\begin{table}[ht]
\centering
\caption{Table of RMSE values for different models}
\begin{tabular}{cc} 
  \hline
  Model & RMSE \\ 
  \hline
  MFCC & 3.420 \\ 
  Chroma & 5.482 \\ 
  Spectral Contrast & 2.783 \\ 
  Temporal & 4.212 \\ 
  Absolute Difference (baseline) & 3.723e+14 \\ 
  \hline
\end{tabular}
\label{tab:rmse-models} 
\end{table}

\section{Conclusion}

In terms of the RMSE evaluation metric, the model based on Spectral Contrast had the lowest error. Spectral contrast provides a detailed view of the spectral peaks and valleys across different frequency bands. It effectively captures the texture and tonal quality differences between the original and the cover, which might be closely related to the emotional or sentiment changes perceived in music. Its superior performance suggests that spectral characteristics like peaks and valleys are essential in capturing nuanced sentiment in music. MFCCs are proficient in capturing the timbral aspects of audio, which include the tone quality or color of a sound. They are particularly good at encapsulating the characteristics of human voice and musical instruments. Their relatively good performance might indicate that timbral elements play a significant role in the sentiment of a song, though not as critical as the spectral contrast. Temporal features, such as the Zero-Crossing Rate and Temporal Centroid, primarily capture rhythm and timing. While these are important characteristics, they might not directly translate to the sentiment of a song, hence the comparatively higher RMSE. Chroma features focus on the harmonic and melodic content of music. While harmonies and melodies are vital in music, their direct influence on sentiment might be more nuanced and not as straightforwardly captured compared to spectral features.

\section{Future Work}

Beyond scaling data collection to increase training examples, future efforts could explore hybrid models that combine all four features to enhance predictive accuracy. Additionally, neural network-based models incorporating video data could capture richer contextual cues. Finally, the framework could be expanded to assess sentiment across a broader range of video types, beyond just music covers.

\vspace{250pt}

%



%
%






\end{document}